\documentclass[useAMS,usenatbib]{mn2e}
\usepackage{epsfig}
\usepackage{amssymb}

\newcommand\ion[2]{#1\,{\scshape{#2}}}%                       % ion, i.e., CII = \ion{C}{ii}

\title[Location of dust in AGNs] {The location of the dust causing internal reddening of active galactic nuclei}
\author[C. Z. P. Heard and C. M. Gaskell]{Clio Z. P. Heard and C. Martin Gaskell\thanks{E-mail:
mgaskell@ucsc.edu}\\
\\Department of Astronomy and Astrophysics, University of California, Santa Cruz, CA 95064
}

\begin{document}

\date{}

\pagerange{\pageref{firstpage}--\pageref{lastpage}} \pubyear{2016}

\maketitle

\label{firstpage}

\begin{abstract}

% Include information about the sample used.

We use the Balmer decrements of the broad-line regions (BLRs) and narrow-line regions (NLRs) of active galactic nuclei (AGNs) as reddening indicators to investigate the location of the dust for four samples of AGNs with reliable estimates of the NLR contribution to the Balmer lines. Intercomparison of the NLR and BLR Balmer decrements indicates that the reddening of the NLR sets a lower limit to the reddening of the BLR. Almost no objects have high NLR reddening but low BLR reddening. The reddening of the BLR is often substantially greater than the reddening of the NLR. The BLR reddening is correlated with the equivalent widths of [\ion{O}{iii}] lines and the intensity of the [\ion{O}{iii}] lines relative to broad H$\beta$. We find these relationships to be consistent with the predictions of a simple model where the additional dust reddening the BLR is interior to the NLR. We thus conclude that the dust causing the additional reddening of the accretion disc and BLR is mostly located at a smaller radius than the NLR.

\end{abstract}


\section{Introduction}

The standard model for active galactic nuclei (AGNs) assumes that the central supermassive black hole is surrounded by an accretion disc (AD) and broad-line region (BLR). These in turn are surrounded by a dusty torus, and a narrow-line region (NLR) is further out. Early optical spectroscopy of active galactic nuclei (AGNs) indicated that there were two broad categories: Seyfert 1s, where broad Balmer lines are prominent in spectra, and Seyfert 2s, where broad Balmer lines are lacking \citep{Khachikian+Weedman71}. The discovery by \citet{Keel80} that Seyfert 1s are preferentially seen face-on led to the unified model for thermal AGNs: that they are all the same type of object, only seen from different viewing angles or at different levels of activity (see \citealt{Antonucci93, Antonucci12}).

Even though it has been recognized since \citet{Keel80} that obscuring off-axis dust is an essential component of AGN models, there has been long-standing controversy over the distribution of this dust. The optically-thick dust in a plane perpendicular to the axis of symmetry of an AGN is commonly considered to be a torus (see \citealt{Antonucci93}) but an alternative possibility is that the obscuring dust needed for the unified model is the outer regions of a wind where dust uplifted from the surface of the outer accretion disc is embedded in an outflow \citep{Konigl+Kartje94}. Closely related to the question of the distribution of dust is the question of how much extinction there is in AGNs (see \citealt{Gaskell15} for a review of the history). The extinction of the various components of an AGN gives us valuable information about the location of the dust.

Since the earliest spectrometry AGNs there has been no doubt that the NLR is reddened  \citep{Wampler68}. \citet{Gaskell82,Gaskell84} and \citet{Wysota+Gaskell88} show that there is agreement between a variety of reddening indicators for permitted and forbidden lines. The median reddening of the NLRs considered by \citet{Gaskell84}  and Wysota \& Gaskell is $E(B-V) \approx 0.30$ with a standard deviation of $\pm 0.35$. From the narrow-line Balmer decrement alone, \citet{Koski78} found a median reddening for Seyfert 2's of 0.52 with a standard deviation of 0.25.

From the {\it Hubble Space Telescope} imaging survey of \citet{Schmitt+03}, the median radius enclosing half the power of the [\ion{O}{iii}] emission is 520 parsecs. However, temporal variability shows that there is substantial [\ion{O}{iii}] emission much closer to the center. For example, \citet{Zheng+95} find the size of the NLR of 3C~390.3 to be $\sim 1$ light-year and \citet{Peterson+13} find the size of the NLR of NGC~5548 to be 3--10 light-years.\footnote{As these authors note, the distances could be larger if the NLR is elongated towards our line of sight.} In contrast to this, the broad-line region (BLR), has a much smaller size  (light-days to light-weeks for a typical Seyfert galaxy), and is located close to the accretion disc (see \citealt{Gaskell09}).

In addition to off-axis dust in the torus, which does not obscure our line of sight for type-1 viewing angles, dust can be located in the following three locations:

\vspace{3mm}

\noindent (A) Between the observer and all components of the AGN. This includes dust in the Milky Way and far from the AGN in the host galaxy.

\noindent (B) Intermingled with the NLR gas.

\noindent (C) Between the NLR and the BLR. This dust might be dust from the torus or in a dusty outflow.

\vspace{3mm}

Our view of the various components of an AGN can be reddened by combinations of dust in different locations. Figure 1 shows the possible configurations of dust in these three locations, and Figure 2 illustrates some of the ways each configuration could occur for a real AGN.

An important question is, might the same dust that causes reddening in the NLR also cause reddening in the BLR and continuum? If the NLR and the BLR are just both reddened by the same foreground dust (See case IV in Figure 2), the reddening should be similar and the reddenings should be correlated. If the dust is predominantly closer to the center than the NLR (e.g., cases III or VII in Figure 2), the BLR reddening will be greater and not correlated with the NLR reddening.  Thus, by comparing the reddening of the NLR and the BLR, we can get a better idea of where the dust is located. In this paper, we compare the reddenings of the NLR and the BLR along with other properties in order to distinguish between possible locations of the dust.

Reddenings can be estimated from observations of line ratios if the theoretical ratio is known. Line ratios used include hydrogen and helium recombination lines and, for low-density gas, some forbidden-line ratios. Permitted lines of \ion{H}{i} and \ion{He}{ii} are produced by both high- and low-density gas. However, the line ratios can potentially be modified by optical depth and collisional effects (see \citealt{Gaskell15} for details). If these effects are significant, they would be greatest for hydrogen. Permitted \ion{He}{ii} lines are affected less because the energy levels are four times higher, the abundance of helium is one tenth of hydrogen, and the size of the He$^{++}$ zone is relatively small. \ion{He}{ii} line ratios are considered to be a good reddening indicator including for the high-density gas of the BLR \citep{Shuder+MacAlpine79,MacAlpine81}.

Although radiative-transfer and collisional effects have widely been considered to be more important for the BLR, a study of blue AGNs (defined as having a continuum slope $\alpha_{\lambda} $ $\gtrsim 1.5$, where $\alpha_{\lambda} $  is fitted in the rest-wavelength range of $4030-5600$ \AA), which are assumed to have low reddening, by \citet{Dong+08} implies that the H$\alpha$/H$\beta$ is close to the theoretical Case B ratio and can be used as a reddening indicator. Investigation of the Balmer decrement of even bluer AGNs implies a mean unreddened, velocity-integrated H$\alpha$/H$\beta$ ratio of $2.72 \pm 0.04$ \citep{Gaskell15}, i.e.\@, within the range of Case B values (see \citealt{Storey+Hummer95}).

For the low-density NLR, calculations have long indicated that the hydrogen-line ratios can be used as a reddening indicator (e.g.\@, \citealt{Gaskell+Ferland84}). This is supported by comparisons between reddenings estimated from NLR Balmer decrements and estimates from other line ratios such as \ion{He}{ii} and forbidden lines (see above). However, because the hydrogen lines are strong, and thus can be measured in a large number of objects, we will only consider reddenings given by the BLR and NLR Balmer decrements in this study.

Dust along the line of sight not only makes things redder, but also makes them fainter. If the dust is predominantly located between the BLR and the NLR, it will make both the accretion disc and the BLR look fainter relative to the NLR. We will therefore also compare the BLR H$\alpha$/H$\beta$ ratio to the [\ion{O}{iii}] equivalent width and to the ratio of [\ion{O}{iii}] to H$\beta_{\mathrm{BLR}}$.

\section{Sample Selection}

In order to analyze the reddening of the NLR and BLR, we chose samples with high-quality spectra. Although a very large number of AGN spectra are available from the SDSS, the quality of many spectra is too low to allow a reliable separation of broad- and narrow-line components, and thus they cannot provide reliable estimates of NLR contribution. Samples from Lick observatory (\citealt{Cohen83} and \citealt{Osterbrock77}) had very high signal-to-noise ratios and careful separation of broad- and narrow-line components. \citet{Dong+05} present a sample of AGNs with very steep Balmer decrements for which they have carefully subtracted the stellar continuum and separated broad- and narrow-line components. We have also included the very blue AGN sample of \citet{Dong+08} which will have very low reddening of the continuum.

%Figure 1
\begin{figure}
 %\vspace{202pt}
 \centering \includegraphics[width = 7cm]{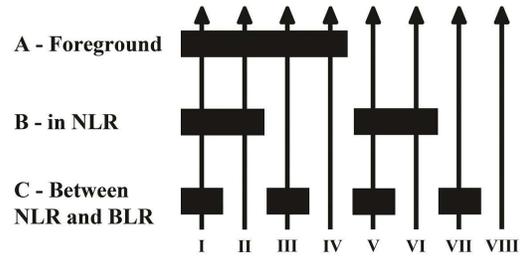}
 \caption{A schematic showing possible combinations of locations of the dust. The vertical arrows represent lines of sight from the observer to the AGN, showing where dust could be found along the line of sight.}
\end{figure}

%Figure 2
\begin{figure}
 %\vspace{202pt}
 \centering \includegraphics[width = 7cm]{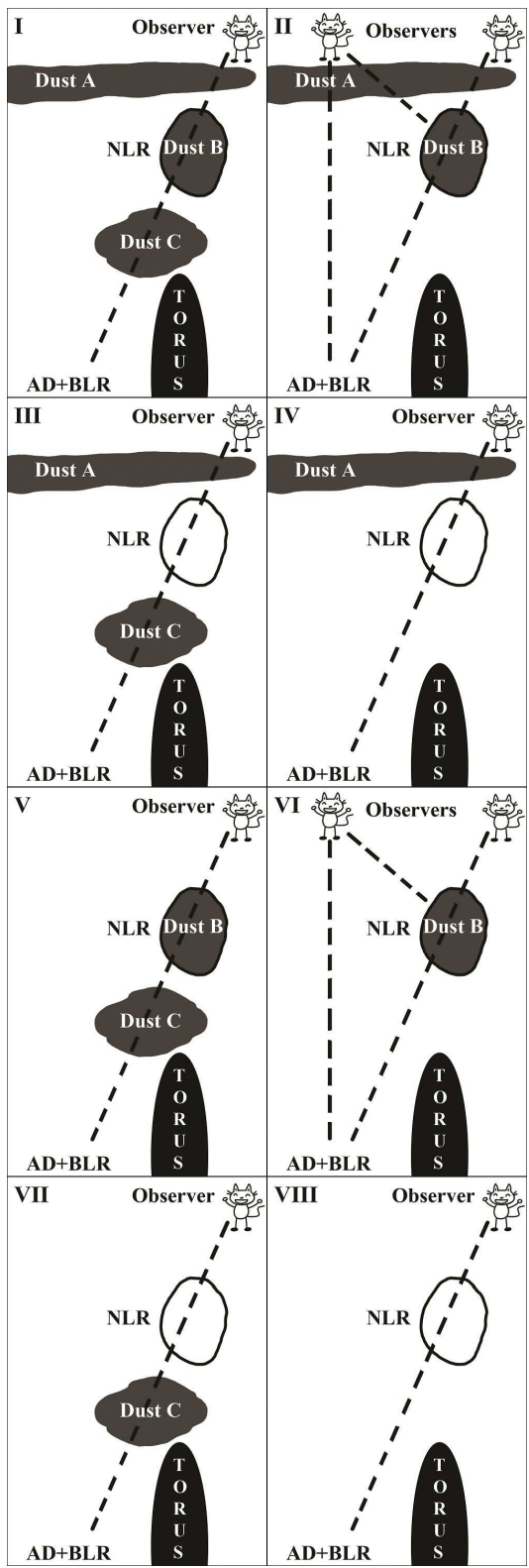}
 %\vspace{100pt}
 \caption{Cartoons illustrating some possible locations for dust along the line of sight of observers to the NLR, BLR, and accretion disc (AD). The roman numerals correspond to the cases in Figure 1 for the observer shown in the upper right of each panel. An additional observer has been added on the upper left in panels II and VI to illustrate how it could be possible to see greater reddening of the NLR than of the BLR and continuum.}
\end{figure}

% section 2
\section{Broad-line versus Narrow-line Reddening}

A straightforward question is whether the reddenings of the BLR and NLR are correlated. If both are reddened by a common screen of dust along the line of sight (case IV in Figures 1 and 2), we would expect their reddenings to be correlated. In Figure 3 we show the broad- and narrow-line Balmer decrements for the 300 highest quality AGN spectra of the sample of \citet{Dong+05} (see their Figure 8). The solid square indicates the expected unreddened line ratios. For the BLR, \citet{Gaskell15} finds H$\alpha$/H$\beta=2.74$. For the NLR, simple recombination theory predicts H$\alpha$/H$\beta \approx 2.9$ (see \citealt{Osterbrock+Ferland06}) and comparison with other line ratios suggests H$\alpha$/H$\beta \approx 3.1 $\citep{Gaskell82, Gaskell84, Wysota+Gaskell88}\footnote{A slightly greater intrinsic H$\alpha$/H$\beta$ ratio for the NLR is expected on theoretical grounds because (a) at a given temperature the Case B ratio for pure hydrogen is lower at high densities (see Figure 4 of \citealt{Gaskell15}), and (b) the temperature is somewhat lower in the NLR than the BLR for the same ionization conditions because there is additional forbidden-line cooling at NLR densities. Lower temperatures cause steeper Balmer decrements (see \citealt{Osterbrock+Ferland06}).}. The arrow indicates the expected trend if the NLR and BLR have common reddening.

Four things can be noticed from Figure 3:
\begin{enumerate}
\item The most noticable thing is,  \textit{at any given NLR Balmer decrement, the flattest BLR decrements (lowest-ratio) are similar to the narrow-line decrement}; i.e., the lower envelope is a 45-degree line parallel to the reddening vector. The null hypothesis, that the minimum BLR decrement is independent of the NLR decrement (see, for example, Figure 2, panel VI), is excluded at a probability $p=0.002$.
\item There is a lack of AGNs with NLR decrements steeper than the BLR decrement. Points to the right of the 45 degree line are consistent with measuring errors. Only one point lies more than two standard deviations to the lower right of the red line.
\item For the majority of AGNs the BLR decrement is steeper than the NLR decrement.
\item There is no significant difference in the median NLR decrement for the AGNs with very steep BLR Balmer decrements compared with those with very flat BLR Balmer decrements.
\end{enumerate}

These results are consistent with the earlier findings of \citet{DeZotti+Gaskell85} and \citet{Glikman+07}. \citet{DeZotti+Gaskell85} found that Balmer decrements for nearby AGNs were steeper for the broad lines than the narrow lines, and \citet{Glikman+07} found that, for the FIRST-2MASS red quasar survey, broad lines seem to be reddened more than the sum of broad and narrow lines.

Our results strongly imply that:
\begin{enumerate}
\item The dust that reddens the NLR also reddens the BLR, but
\item most of the dust reddening the BLR lies between the BLR and the NLR.
\end{enumerate}

The dust causing the common reddening of both the NLR and BLR could be associated with the NLR or it could be dust in the host galaxy. \citet{DeZotti+Gaskell85} studied nuclear continuum colors and both broad- and narrow-line Balmer decrements of nearby AGNs as a function of the orientation of the host galaxy. They found that for more highly inclined disc galaxies the nuclear colors were redder and the Balmer decrements steeper. Since the axis of symmetry of an AGN is not necessarily aligned with the rotation axis of the galaxy (see, for example, \citealt{Tohline+Osterbrock82, Valtonen83, Battye+Browne09}), the greater nuclear reddening of inclined galaxies must be due in part simply to dust in the plane of the galaxy.

 %Figure 3
\begin{figure}
 \centering \includegraphics[width=8.5cm]{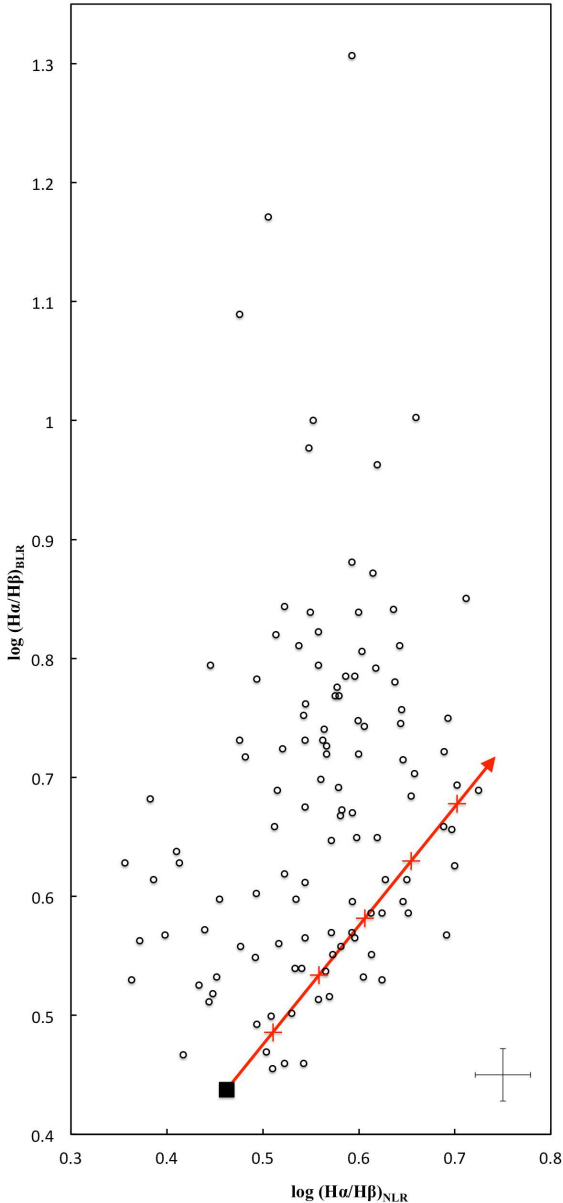}
 \caption{Intercomparison of the H$\alpha$/H$\beta$ ratios for the NLR and BLR. The black square corresponds to unreddened Case B values of H$\alpha$/H$\beta_{\mathrm{NLR}}=2.9$, H$\alpha$/H$\beta_{\mathrm{BLR}}=2.74$. The arrow corresponds to common reddening of the NLR and BLR. The crosses along the reddening vector indicate reddening increments of $\Delta E(B-V)=0.1$. A typical error bar is shown at the lower right. Data from  \citet{Dong+05}. Lower limits to H$\alpha$/H$\beta_{\mathrm{BLR}}$ are not included.}

\end{figure}

% section 4
\section{The Relative Strength of [\ion{O}{iii}] versus the BLR Reddening}

As noted above, intervening dust not only causes reddening, but also extinction. If the dust is located between the NLR and the BLR, we would expect higher extinction of the BLR and accretion disc to make [\ion{O}{iii}] stronger relative to the BLR and the continuum.

% section 4.1
\subsection{Selection of Objects}

[\ion{O}{iii}] $\lambda$5007 is the strongest narrow line in the optical and the Balmer lines are the strongest broad lines. Therefore, a very large number of measurements are potentially available in the Sloan Digital Sky Survey (SDSS). \citet{Dong+08} have carefully measured H$\alpha$/H$\beta$ ratios after subtraction of narrow lines and \ion{Fe}{ii}, etc. Comparison of their ratios with the simple gaussian fits of the SDSS pipeline reveals quite a few very discrepant measurements in the SDSS database and that the difference between the \citet{Dong+08} measurements and the SDSS is around $\pm 0.2$ dex (i.e., a factor of 50\%).

The Lick Observatory data of \citet{Osterbrock77} and \citet{Cohen83} are of much higher signal-to-noise ratio than most SDSS spectra. This allowed them to carry out careful deblending of lines. We have therefore used the Lick data and the line intensity measurements  of \citet{Dong+05,Dong+08} from SDSS spectra. The latter do not give [\ion{O}{iii}] intensities and H$\beta$ equivalent widths, so we have taken these values from the SDSS.

% section 4.2
\subsection{[\ion{O}{III}]/H$\beta$}

Figure 4 shows the [\ion{O}{iii}]/H$\beta_{\mathrm{BLR}}$ ratio versus the broad line H$\alpha$/H$\beta$ ratio. The red lines indicate the expected change in the ratios for reddening of the BLR alone. The separation between the two lines indicates the probable intrinsic scatter in the [\ion{O}{iii}]/H$\beta_{\mathrm{BLR}}$ ratio. It can be seen that there is a positive correlation between the [\ion{O}{iii}]/H$\beta_{\mathrm{BLR}}$ ratio and the BLR Balmer decrement, suggesting that dust lying between the two emitting regions causes extinction of the continuum.  We note that the SDSS observations show more scatter than the higher quality Lick observatory observations.

The H$\beta$ flux appears in the denominator of both axes in Figure 4, so an error in H$\beta$ could cause a spurious correlation. However, this correlation, indicated by the green arrow in Figure 4, is mostly not in the same direction as the observed correlation and the effect is expected to be small. The length of the arrow corresponds to a 20\% error in the H$\beta$ flux (more than double the typical error of \citealt{Dong+08}). Nonetheless, it is possible that large measuring errors for very weak H$\beta$ lines enhanced the correlation in the upper right of the diagram. To test for this, in Figure 5 we show the  [\ion{O}{iii}]/H$\alpha_{\mathrm{BLR}}$ ratio versus H$\alpha$/H$\beta$. In this case large measuring errors would cause a decrease in log[\ion{O}{iii}]/H$\alpha_{\mathrm{BLR}}$ as indicated by the green arrow. The similarity between Figures 4 and 5 argues against the trend in Figure 4 being due to measuring errors in the H$\beta$ flux.

The observations of \citet{Dong+05} of AGNs with very steep Balmer Decrements  (indicated by diamond symbols in Figures 4 and 5) seem to have systematically lower [\ion{O}{iii}]/H$\beta_{\mathrm{BLR}}$ values. We suspect that the lower signal noise of the SDSS spectra could be contributing to this as well. Visual inspection of the spectra suggests that the discrepancy is due to underestimation of the flux in the [\ion{O}{iii}] line broad wings, which contain a substantial amount of the [\ion{O}{iii}] flux. There is probably also a selection bias against finding AGNs with extremely high [\ion{O}{iii}]/H$\beta_{\mathrm{BLR}}$ ratios because broad H$\beta$ is so weak. Objects with very high [\ion{O}{iii}]/H$\beta_{\mathrm{BLR}}$ would be classified as Seyfert 2s.

%Figure 4
\begin{figure}
 \centering \includegraphics[width=8.5cm]{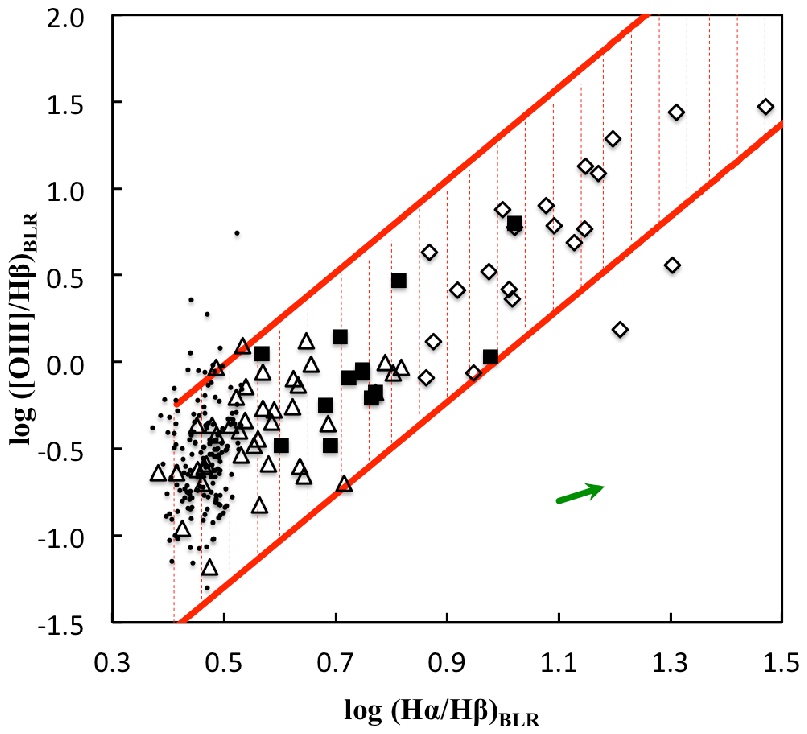}
 \caption{Ratio of  [\ion{O}{iii}]/H$\beta_{\mathrm{BLR}}$ plotted against the broad line H$\alpha$/H$\beta$ ratio. Triangles are measurements from \citet{Osterbrock77}, small dots are from \citet{Dong+08} and the SDSS, diamonds are measurements from \citet{Dong+05} and the SDSS, and squares are measurements from \citet{Cohen83}. The  lines show the expected variation in [\ion{O}{iii}]/H$\beta$ for unreddened ratios of 0.002 (bottom line) and 0.019 (top line). The vertical dashed lines indicate $\Delta E(B-V)$ of 0.1 The green arrow shows the effect of a 20\% error in the H$\beta$ flux.}
 \end{figure}

The correlations in Figures 4 and 5 and the qualitative agreement with a simple BLR reddening model are consistent with reddening of AGNs mostly being caused by dust between the NLR and BLR.

% Section 4.3
\subsection{Apparent Equivalent Width of [\ion{O}{III}]}

 As an additional check we also looked at the ratio of [\ion{O}{iii}] to the continuum, i.e., the equivalent width (EW) of [\ion{O}{iii}]. The observed continuum includes light from the AGN and starlight in the host galaxy. Figure 6 shows the EW of [\ion{O}{iii}] against the broad line H$\alpha$/H$\beta$ ratio. \citet{Cohen83} did not give EWs of [\ion{O}{iii}], so those results (the squares in Figures 4 and 5) are not included. For the flatter BLR Balmer decrements, there is the expected increase in apparent equivalent width of [\ion{O}{iii}]. However, for steeper decrements, the apparent equivalent width appears to plateau. Again, there is more scatter for the SDSS measurements and, for reasons that have been discussed above, the [\ion{O}{iii}] equivalent widths are systematically lower.

%figure 5
\begin{figure}
 %\vspace{202pt}
 \centering \includegraphics[width=8.5cm]{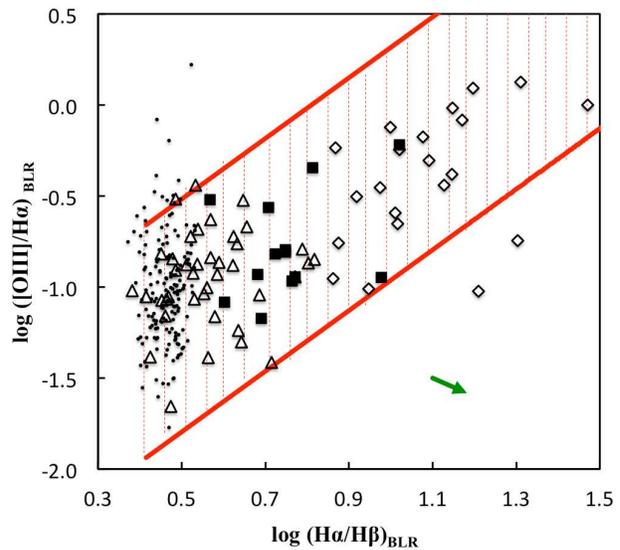}
 \caption{As for Figure 4, but showing the ratio of [\ion{O}{iii}] to H$\alpha$. The green arrow shows the effect of a 20\% measuring error in the flux of H$\alpha$.}
\end{figure}

%figure 6
\begin{figure}
 %\vspace{202pt}
 \centering \includegraphics[width=8.5cm]{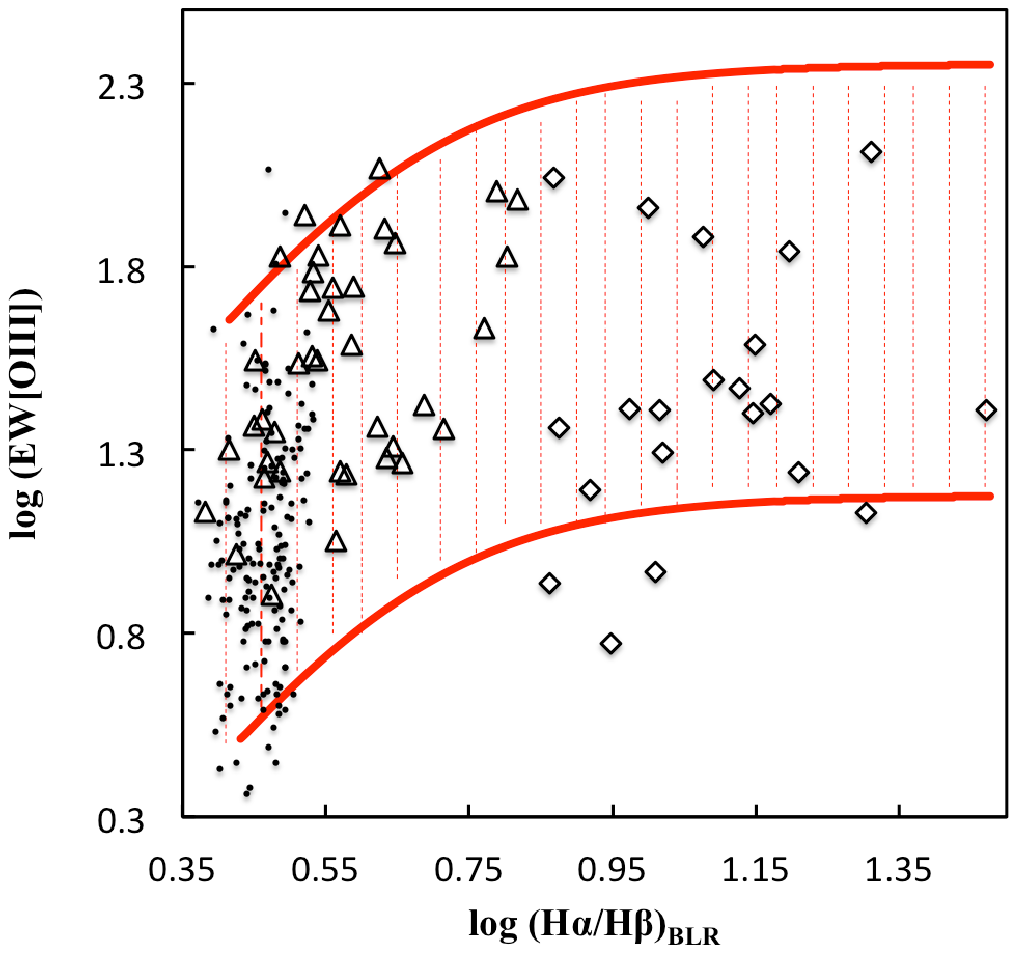}
 \caption{The logarithms of the equivalent widths of [\ion{O}{iii}] $\lambda$5007 (in \AA) plotted against the broad line H$\alpha$/H$\beta$ ratio. Triangles are measurements from \citet{Osterbrock77}, diamonds are measurements from \citet{Dong+05} and the SDSS, and small dots are from \citet{Dong+08} and the SDSS. The curves show the changes in apparent equivalent width (relative to both the accretion disc {\em and} the host galaxy) predicted by our simple model as discussed in the text. The lower curve corresponds to an unreddened apparent equivalent width ratio 3\AA \@  and the upper curve to 45\AA. The fraction of host galaxy light at $\lambda$5000 has been taken to be 0.2. The vertical dashed lines indicate $\Delta E(B-V)$ of 0.1.}
\end{figure}

As shown by the curves in Figure 6, the plateauing in the apparent equivalent width of [\ion{O}{iii}] is just what is expected from the simple BLR reddening model. This is because observed EWs measure the intensity of an emission line relative both to the light of the accretion disc and the starlight of the host galaxy. We assume that the reddening only affects the light of the inner regions of the AGN and not the starlight. This naturally produces the saturation in [\ion{O}{iii}] equivalent width. To give the approximate envelopes of the distribution in Figure 6, the unreddened EW varies by $\pm 0.4$ dex and the average fraction of host galaxy starlight for the least-reddened AGNs is $\approx 0.2$ at $\lambda$5000.

It is interesting to compare the contributions of host galaxy starlight needed to explain the variation of [\ion{O}{iii}] equivalent width with independent estimates of host galaxy fractions. The reddening of most BLRs and accretion discs is not zero but it is easy to predict the fractions of host galaxy starlight as a function of H$\alpha$/H$\beta$ from our model. The AGNs in the Osterbrock sample include many well-studied bright AGNs. The median H$\alpha$/H$\beta \approx4$. For this value, our model described in the previous paragraph predicts a galaxy starlight fraction of 0.4. {\it Hubble Space Telescope} imaging by \citet{Bentz+13} of reverberation-mapped AGNs (a sample with large overlap with \citealt{Osterbrock77} and observed with similar ground-based apertures) shows a mean fraction of 0.42.

\citet{VandenBerk+06} have estimated the fractional contributions of the host galaxy starlight at $\lambda$4200 for most of the SDSS AGNs considered here. The median host galaxy starlight fraction at $\lambda$4200 for the \citet{Dong+08} objects is 0.08 while the median host galaxy fraction for the \citet{Dong+05} objects is 0.6.

The consistency between the variation of the equivalent width of [\ion{O}{iii}] with reddening and the simple predictions (Figure 6) further reinforces the picture that the dust causing reddening in AGNs lies mainly between the NLR and the BLR.

\section{Conclusions}

Using observed BLR Balmer decrements as reddening indicators, we conclude from the sample of AGNs we have studied that

\begin{enumerate}

\item The BLR is almost always reddened at least as much as the NLR, suggesting there is common intervening reddening in most cases.

\item As the reddening of the BLR increases, the relative intensity of the [\ion{O}{iii}] lines increases compared to both broad H$\beta$ and to the continuum.

\item This increase in [\ion{O}{iii}]/H$\beta_{\mathrm{BLR}}$ and apparent EW([\ion{O}{iii}]) with increasing BLR reddening is naturally explained by additional reddening of the BLR and continuum alone.

If these results apply to all AGNs, they imply two important things. Firstly, they provide support for steep BLR decrements being a consequence of reddening rather than intrinsic steepening due to radiative transfer and collisional effects in the BLR clouds. Secondly, they imply that the dust causing reddening of the continuum and BLR is mostly located between the NLR and the BLR. The most natural origin of this dust is to be blown off of the torus, or indeed even to be what is regarded as the torus. The common reddening of the NLR and BLR could be due to dust associated with the NLR or other dust in the host galaxy.

\end{enumerate}

\section*{Acknowledgments}

We are grateful for the feedback and advice from Xiaobo Dong and Ski Antonucci. We thank the anonymous referee for detailed and helpful comments which have improved the paper.

%\label{lastpage}

\end{document}